# Design in Complex Systems: Individual Performance versus System Efficiency

Chengling Gou

Physics Department, Beijing University of Aeronautics and Astronautics
37 Xueyuan Road, Haidian District, Beijing, China, 100083

Physics Department, University of Oxford
Clarendon Laboratory, Parks Road, Oxford, OX1 3PU, UK
gouchengling@hotmail.com, c.gou1@physics.ox.ac.uk

Abstract: This paper studies the correlations of the average winnings of agents and the volatilities of systems based on mix-game model which is an extension of minority game (MG). In mix-game, there are two groups of agents; group1 plays the majority game, but the group2 plays the minority game. The results show that the correlations between the average winnings of agents and the mean of local volatilities are different with different combinations of agents' memory lengths when the proportion of agents in group1 increases. The average winnings of agents can represent the average individual performance and the volatility of a system can represent the efficiency of the system. Therefore, these results imply that the combinations of agents' memory lengths largely influence the relation between the system efficiency and the average individual performance. These results could give some guidance in designing complex systems.

1. **Introduction:**

   Systems consisting of heterogeneous agents who compete constantly for limited resources, such as different species in an ecological system, traders in financial markets, drivers on city roads and highways or users accessing a computer network – raise problems of coordination. All agents in such systems want to optimize their performances by making decisions based on available information and their strategies, and individual goals will most certainly be conflicting. A designer or a manager outside such a system may hope to balance the individual performance and the efficiency of the system. He or she need to know the relationship between the individual performance and the efficiency of the system in order to reach his or her goal. However, the interplay between the state of optimal individual performance and the reach of a globally efficient phase where the collective use of resources is optimal may be quite subtle [1]. Therefore, the study of this issue has both theoretical and practical importance, which can reveal how the macroscopic properties of a system may emerge from the microscopic interaction of agents and under which conditions, and this understanding may help for designing of complex systems.

Agent-based models can be used to address directly the relation between microscopic behavior of agents and macroscopic properties of a system (like fluctuations, predictability and efficiency) [1]. The average winnings of agents can represent the average individual performance and the volatility of a system can represent the efficiency of the system. Therefore, in this paper, I study this issue by examining the correlations of the average winnings of agents and the volatilities of systems based on mix-game model [2] which is an extension of minority game (MG) [3, 4, 5].

The structure of this paper is as following. Section 2 describes the mix-game model and the simulation conditions. Section 3 presents the simulation results, the calculation results of the quantitative correlations and relevant discussion. Section 4 gives some suggestions for potential applications of the results. In section 5, the conclusion is reached.

2. **Mix-game model:**

The MG comprises an odd number of agents choosing repeatedly between the options of buying (1) and selling (0) a quantity of a risky asset. The agents continually try to make the minority decision i.e. buy assets when the majority of other agents are selling and sell when the majority of other agents are buying. Mix-game model is an extension of MG, so its structure is similar to MG. In mix-game, there are two groups of agents; group1 plays the majority game [6, 7, 8, 9, 10, 11, 12, 13], but the group2 plays the minority game. N (odd number) is the total number of the agents and N1 is number of agents in group1. The system resource is r = N*L, where L<1 is the proportion of resource of the system. All agents compete in the system for the limited resource r. T1 and T2 are the time horizon lengths of the two groups, and m1 and m2 denote the memory lengths of the two groups, respectively.

The global information only available to the agents is a common bit-string "memory" of the m1 or m2 most recent competition outcomes (1 or 0). A strategy consists of a response, i.e., 0 (sell) or 1 (buy), to each possible bit string; hence there are $2^{2^{m1}}$ or $2^{2^{m2}}$ possible strategies for group1 or group2, respectively, which form full strategy spaces (FSS). At the beginning of the game, each agent is assigned s strategies and keeps them unchangeable during the game. After each turn, agents assign one (virtual) point to a strategy which would have predicted the correct outcome. For agents in group1, they will reward their strategies one point if they are in the majority; otherwise, for agents in group2, they will reward their strategies one point if they are in the minority. Agents collect the virtual points for their strategies over the time horizon T1 or T2, and they use their strategies which have the highest virtual point in each turn. If there are two strategies which have the highest virtual point, agents use coin toss to decide which strategy to be used. Excess demand is equal to the number of ones (buy) which agents choose minus the number of zeros (sell) which agents choose. According to a widely accepted assumption that excess demand exerts a force on the price of the asset and the change of price is proportion to the excess demand in a financial market [14, 15, 16], the time series of price of the asset can be calculated based on the time series of excess demand.

In simulations, the distribution of initial strategies of agents is randomly distributed and keeps unchanged during the games. Simulation turns are 3000. The window length of local volatility is 5. Total number of agents is 201. Number of strategies per agent is 2.

## 3. Results and discussions:

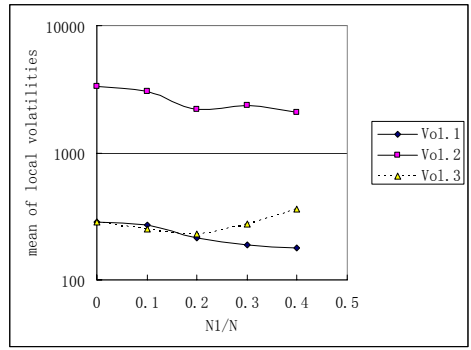

Fig.1 means of local volatilities vs. different N1/N; Vol.1 representing mean of local volatilities of m1=m2=6, T1=T2=60; Vol.2 representing mean of local volatilities of m1=6, m2=3, T1=60, T2=12; Vol.3 representing mean of local volatilities of m1=3, m2=6, T1=12, T2=60.

First I study how the means of local volatilities change when the proportion of agents in group1 increases. Fig.1 shows that means of local volatilities (Vol.1 and Vol.2) decrease while N1/N increases from 0 to 0.4 under condition of m1=m2=6, T1=T2=60 and condition of m1=6, m2=3, T1=60, T2=12, respectively. But under condition of m1=3, m2=6, T1=12, T2=60, the mean of local volatilities (Vol.3) has a minimum value at N1/N=0.2.

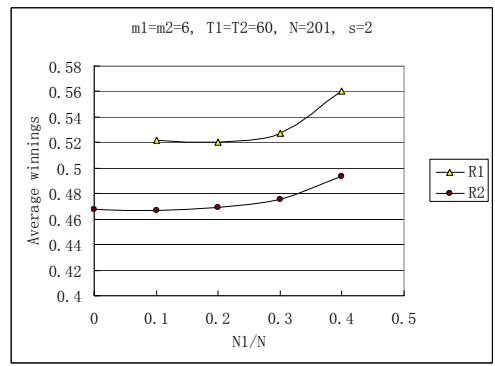

Fig.2 average winnings per agent per turn vs. different proportion of agents in group1 when m1=m2=6, T1=T2=60; R1 represents the average winning per agent per turn in group1, and R2 represents the average winning per agent per turn in group2.

Table 1 correlations of R1, R1 and Vol.1 under the condition of m1=m2=6, T1=T2=60

| Correlation | R1 | R2 | Vol.1 |
|---|---|---|---|
| R1 | 1 | | |
| R2 | 0.98 | 1 | |
| Vol.1 | −0.63 | −0.76 | 1 |

Then I study the average winnings of agents. From Fig.2, one can find that the average winnings (R1 and

R2) of these two groups increase when N1/N is larger than 0.1 and the average winning of group1 is larger than those of group2 under the condition of m1=m2=6, T1=T2=60. Agents in both groups benefit from the increase of the number of agents in group1. Comparing Fig.2 with Fig.1, one can find that the mean of local volatilities (Vol.1) decreases accompanying with the increase of the average winnings of group1 and group2 (R1, R2) while N1/N increases from 0 to 0.4 under this condition. Table 1 show that the correlations between local volatilities and the average winnings of both groups are negative while the correlation of the average winnings of these two groups is positive.

This result means that the improvement of the efficiency of systems is accordant with the improvement of the performance of individual agents under this simulation condition. Both groups and the whole system benefit from the increase of the number of agents in group1. Similar phenomena can be found in ecological systems, computing systems and economic systems in which agents (species, computing tasks and firms) having different niches will improve the efficiencies of systems and their own performance [17].

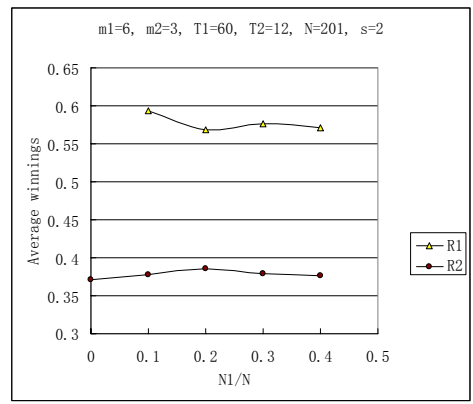

Fig.3 average winnings per agent per turn vs. different proportion of agents in group1 when m1=6, m2=3, T1=60, T2=12. R1, R2 have the same meaning as that indicated in Fig.2.

Table 2 correlations of R1, R1 and Vol.2 under the condition of m1=6, m2=3, T1=60, T2=12

| Correlation | R1 | R2 | Vol.2 |
|:---:|:---:|:---:|:---:|
| R1 | 1 | | |
| R2 | −0.48 | 1 | |
| Vol.2 | 0.98 | −0.67 | 1 |

From Fig. 3, one can find that the average winnings (R1 and R2) of these two groups don't change obviously when N1/N increases from 0 to 0.4, except R1 at N1/N=0.1. The average winnings of group1 are larger than those of group2. Comparing Fig.3 with Fig.1, one can find that the mean of local volatilities (Vol.2) decreases while N1/N increases from 0 to 0.4 under the condition of m1=6, m2=3, T1=60, T2=12, but the change of the average winnings of group1 and group2 (R1, R2) seems more complicated than those in Fig.2. Table 2 gives a clearer picture about the correlations among R1, R2 and Vol.2 under this simulation condition: the correlation between R1 and R2 is negative; the correlation between R1 and Vol.2 is positive while the correlation between R2 and Vol.2 is negative.

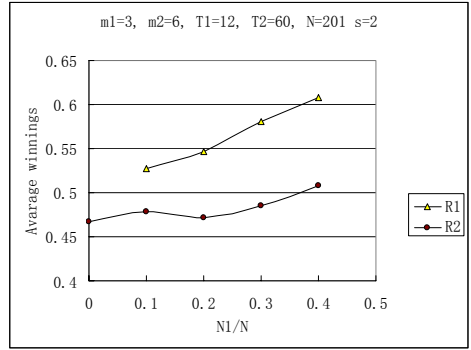

Fig.4 average winnings per agent per turn vs. different proportion of agents in group1 when m1=3, m2=6, T1=12, T2=60. R1, R2 have the same meaning as that indicated in Fig.2.

Table 3 correlations of R1, R1 and Vol.3 under the condition of m1=3, m2=6, T1=12, T2=60,

| Correlation | R1 | R2 | Vol.3 |
|---|---|---|---|
| R1 | 1 | | |
| R2 | 0.87 | 1 | |
| Vol.3 | 0.89 | 0.82 | 1 |

From Fig. 4, one can find that the average winnings (R1 and R2) of these two groups increase obviously when N1/N increases from 0 to 0.4 and the average winning of group1 (R1) is larger than that of group2 (R2). Agents in both groups benefit from the increase of the number of agents in group1. Comparing Fig.4 with Fig.1, one can find that the mean of local volatilities (Vol.3) decreases slightly when N1/N increase from 0 to 0.2, then it increase while N1/N increases from 0.2 to 0.4 under the condition of m1=3, m2=6, T1=12, T2=60, accompanying with the increase of the average winnings of group1 and group2 (R1, R2). Table 3 shows that the correlations between local volatilities and the average winnings of both groups are positive and the correlation of average winnings of these two groups is also positive. This means that the improvement of the performance of individual agents accompanies with the decrease of the system efficiency under this simulation condition. Agents can make profits from the larger fluctuation of systems, which is accordant with the reality of financial markets [18].

4. **Potential applications----an open question**

The above results could give the following suggestions when we use mix-game model to design or model complex systems.
- If we want to design a system with both high efficiency of the system and high individual performance, we need to make the agents have different payoffs, the same memory lengths and a relatively large number of agents in group1.
- If we use mix-game to model financial markets, we need to make m1 smaller than m2.

5. **Concluding remarks:**

In mix-game model, the correlations between the average winnings of agents (R1 and R2) and the mean of local volatilities are different with different combinations of memory lengths of agents (m1 and m2) when the proportion of agents in group1 increases. The combinations of agents' memory lengths largely influence the relation of the system efficiency and the average individual performance. This result could give some guidance in designing complex systems.

**Acknowledgements**

This research is supported by Chinese Scholarship Council. Thanks Professor Neil F. Johnson for suggesting modification of agents' memories and discussing about the calculation of average winning of agents. Thanks David Smith for providing the original program code of MG.